\documentclass[preprint,showkeys,showpacs,amsmath,amssymb,nofootinbib]{revtex4}

\usepackage{bm}

\begin{document}

\title{Acceleration-extended Newton-Hooke symmetry and its dynamical realization}

\author{Fu-Li Liu}\email{fuliliu2008@yahoo.com.cn}
\affiliation{Department of Physics, Beijing Institute of Technology\\
Beijing 100081, P. R. China}
\author{Yu Tian}\email{ytian@gucas.ac.cn}\thanks{Tel:
(86)10-82792321}
\affiliation{College of Physical Sciences, Graduate University of Chinese Academy of Sciences\\
Beijing 100049, P. R. China}

\date{\today}

\begin{abstract}
Newton-Hooke group is the nonrelativistic limit of de Sitter
(anti-de Sitter) group, which can be enlarged with
transformations that describe constant acceleration.
We consider a higher order Lagrangian that is quasi-invariant under the
acceleration-extended Newton-Hooke symmetry, and obtain
the Schr\"{o}dinger equation quantizing the Hamiltonian
corresponding to its first order form. We show that the Schr\"{o}dinger equation
is invariant under the acceleration-extended Newton-Hooke transformations.
We also discuss briefly the exotic conformal Newton-Hooke symmetry
in $2+1$ dimension.
\end{abstract}



\maketitle

\section{Introduction}
The Newton-Hooke (NH) symmetry can be obtained by taking the
nonrelativistic limit of de Sitter (dS) or anti-de Sitter (AdS)
symmetry.\footnote{That is the so-called Newton-Hooke limit
\cite{Huang,Tian}: the speed of light $c\rightarrow\infty$,
cosmological constant $\Lambda\rightarrow 0$, but $c^2\Lambda$ is
kept fixed.} If we define $\frac{c^2\Lambda}{3}=\frac{1}{R^2}$, the
parameter $R$ has the dimension of time. When we take further
contraction $R\rightarrow\infty$, we will obtain the familiar
Galileo symmetry and the corresponding standard Newton space-time.
The NH symmetry and the corresponding NH space-time have interesting
applications in nonrelativistic cosmology \cite{J} and even in
String/M-theory \cite{Gibbons,Gao}, besides the significance in
fundamental physics \cite{Bacry,Huang,Tian}.

Recently, acceleration-extended Galileo symmetry with central
charges and its dynamical realization has been studied in
Ref.\cite{P}; these results have been generalized to the NH
space-time, and it is shown that the acceleration-extended NH
symmetry has one central charge in arbitrary dimension and three
central charges in $2+1$ dimension (the so-called exotic case) in
Ref.\cite{Stichel}. In fact, the exotic central extension of NH
symmetry, with or without acceleration extension, has been
extensively studied in the literature \cite{AGKP,AGKP2,GL}.

Following those results, in this paper, we focus on dynamical
realization of the symmetry, i.e. the nonrelativistic Lagrangian on NH space-time
which is quasi-invariant under the acceleration-extended NH
symmetry, with one central charge. Furthermore, we
obtain the first order form of the Lagrangian by introducing extra
variables, and write down the Sch\"{o}dinger equation of the system
under geometric quantization of the corresponding Hamiltonian. Naturally, we consider
whether the Sch\"{o}dinger equation is invariant under the
acceleration-extended NH transformations, and the answer is
confirmative. In fact, since the (classical) acceleration-extended NH transformations
act on the phase space (extended by extra variables) as nontrivial canonical transformations,
the corresponding quantum ones are nontrivial unitary transformations on the Hilbert space.
We also discuss briefly the exotic conformal NH
algebra (in $2+1$ dimension) and its dynamical realization.

We organize this paper as follows. In Section \ref{sec:algebra}, we recall some known
results on the NH algebra and add to it transformations describing
constant acceleration. We also consider the central element
$\kappa$ there. Then, we construct the classical Lagrangian with
higher order time derivatives which is quasi-invariant under the
acceleration-extended Newton-Hooke symmetry in Section \ref{sec:Lagrangian}. In
Section \ref{sec:1st-order} we obtain the extended phase space formulation
from the first order form of the Lagrangian. We also derive the
Schr\"{o}dinger equation in standard way and show its invariance
under the acceleration-extended (and also central extended) $NH$ transformations.
In the following section we discuss the exotic conformal NH algebra and its
dynamics. The last section contains our conclusion.

\section{acceleration-extended Newton-Hooke algebra with central charge}\label{sec:algebra}
First, let us recall the NH algebra, which is the nonrelativistic
limit of de Sitter algebra.\footnote{The anti-de Sitter case can be dealt with in parallel,
without any difficult.} The nonvanishing commutators are
$(i,j,k,l=1,2,\cdots,D-1)$
\begin{subequations}
\label{eq:whole}
\begin{eqnarray}\label{eq:NH}
&\left[J_{ij},J_{kl}\right]=\delta_{il}J_{jl}-\delta_{jl}J_{ik}+\delta_{jk}J_{il}-\delta_{il}J_{jk}\\
&\left[J_{ij},A_k\right]=\delta_{ik}A_j-\delta_{jk}A_i\qquad\left(A_i=P_i,K_i\right) \\
&\left[H,K_i\right]=P_i\\
&\left[H,P_i\right]=\frac{K_i}{R^2}\label{eq:last}
\end{eqnarray}
\end{subequations}
where $J_{ij}$ are the generators of spatial rotation, $H$
that of time translation, $P_i$ those of spatial translation and
$K_i$ those of boost.

We can make the following central extension in arbitrary dimension $D$:
\begin{eqnarray}
\left[P_i,K_j\right]=-i m\delta_{ij}
\end{eqnarray}
as in the Galileo case, where $m$ describes the nonrelativistic mass appearing in the NH quantum
mechanics \cite{Tian}.

We can also add to NH algebra generators $F_i$ describing
acceleration transformations to enlarge it to the so-called
$\widehat{NH}$ algebra. It is known that the constant acceleration
transformations in NH space-time can be obtained by some appropriate combination of spatial
translations and special conformal transformations \cite{Yu}.
Written in terms of the so-called static coordinates \cite{Tian} in
NH space-time, the acceleration transformation is
\begin{eqnarray}
x_i'=x_i+2R^2 (\cosh\frac{t}{R}-1) b_i
\end{eqnarray}
which is the same as that in the Galileo case, when $R
\rightarrow\infty$. Combining it with the standard NH transformation
\cite{Gao,Tian}, we can have the full $\widehat{NH}$
transformation as follows:
\begin{eqnarray}\label{eq:hat-NH}
&x_i'=v_i R\sinh\frac{t}{R}+a_i\cosh\frac{t}{R}+2R^2 (\cosh\frac{t}{R}-1) b_i+{O_i}^j x_j\nonumber\\
& t'=t+a
\end{eqnarray}
where ${O_i}^j$ is the transformation matrix generated by spatial
rotation $J_{ij}$; $v_i$ is a ``velocity" corresponding to boost
$K_i$; $a_i$ and $b_i$ are spatial translations generated
by momentum operators $P_i$ and ``accelerations" generated by
acceleration operators $F_i$, respectively. Of course, in the
space-time realization of the NH (or $\widehat{NH}$) symmetry, there
is no room for the central parameter $m$, i.e. we must have
$\left[P_i,K_j\right]=0$.

The generators of transformation (\ref{eq:hat-NH}) can be
represented by tangent vector fields on the NH space-time. By
definition, the generators of spatial translations are described by
the vector fields
\begin{eqnarray}
P_i=\left.\frac{\partial x_j'}{\partial
a_i}\right|_{a=b=v=0,O=1}\frac{\partial}{\partial x_i}=\cosh\frac
{t}{R}\frac{\partial}{\partial x_i}
\end{eqnarray}
Similarly, we can write down the vector fields of time translation
$H$, boost $K_i$, spatial rotation $J_{ij}$ as well as acceleration
generators $F_i$ as follows:
\begin{eqnarray}
&H=\frac{\partial}{\partial t},\qquad
K_i=R\sinh \frac{t}{R}\cdot\frac{\partial}{\partial x_i}\nonumber\\
&J_{ij}=x_i\frac{\partial}{\partial x_j}-x_j\frac{\partial}{\partial
x_i},\qquad
F_i=2R^2\left(\cosh\frac{t}{R}-1\right)\cdot\frac{\partial}{\partial
x_i}
\end{eqnarray}

One can check that the above form of generators satisfy the
relations (\ref{eq:NH}-\ref{eq:last}) and
\begin{eqnarray}\label{eq:acceleration}
\left[J_{ij}, F_k\right]=\delta_{ik}F_j-\delta_{jk}F_i,\qquad [H,
F_i]=2K_i
\end{eqnarray}
The other commutators of $F_i$ are vanishing. In fact, without
considering the space-time realization, one can show that the Jacobi
identity for generators $(H,P_i,F_j)$ implies $m=0$ if $[K_i,
F_j]=0$. Just relaxing this condition, in arbitrary space-time
dimension $D$, we can introduce at least one central element
$\kappa$ that appears as (see \cite{Stichel})
\begin{eqnarray}
[K_i, F_j]=2\kappa\delta_{ij}
\end{eqnarray}
Using the Jacobi identities, we can obtain
\begin{eqnarray}
[P_i, K_j]=-\frac{\kappa}{R^2}\delta_{ij}
\end{eqnarray}
with other commutators kept unchanged.

\section{Lagrangian with acceleration-extended Newton-Hooke (quasi-)symmetry in arbitrary dimension $D$}
\label{sec:Lagrangian}
The finite $\widehat{NH}$ transformation (\ref{eq:hat-NH}) gives the
infinitesimal transformation as follows:
\begin{eqnarray}\label{eq:infinitesimal}
&\delta x_i={(\delta O)_i}^j x_j+\cosh\frac{t}{R}\delta
a_i+R\sinh\frac{t}{R}\delta v_i+2R^2(\cosh\frac{t}{R}-1)\delta
b_i\nonumber\\
&\delta t=\delta a
\end{eqnarray}
where $\delta a_i$, $\delta b_i$, $\delta a$ and ${(\delta
O)_i}^j$ are infinitesimal parameters corresponding to the finite
ones in (\ref{eq:hat-NH}).

Then, we should look for a Lagrangian which is
quasi-invariant\footnote{That, following Ref.\cite{P}, means that
the Lagrangian is invariant up to a boundary term of the form
$\frac{d}{dt}f(x_i,\dot{x}_i)$, in comparison with the invariant one
with usual $\frac{d}{dt}f(x_i)$.} under transformations
(\ref{eq:infinitesimal}) and which, in the limit $R\rightarrow
\infty$, goes back to the Lagrangian with higher order time
derivatives in Galileo case (see \cite{P}), namely:
\begin{eqnarray}
L|_{R\rightarrow \infty}=\frac{\kappa}{2}{\ddot x_i}^2
\end{eqnarray}
where an over dot means $\frac{d}{dt}$ as usual.

The extension of this Lagrangian to the NH case can be obtained
through the following substitution \cite{Stichel}:
\begin{eqnarray}
{\ddot x_i}^2 \rightarrow {\ddot x_i}^2+\frac {1}{R^2}{\dot{x_i}}^2
\end{eqnarray}
 Then, we can obtain the Lagrangian in NH case :
\begin{eqnarray}\label{eq:Lagrangian}
L=\frac{\kappa}{2}({\ddot x_i}^2+\frac{1}{R^2}{\dot{x_i}}^2)
\end{eqnarray}
It is easy to check that the Lagrangian
(\ref{eq:Lagrangian}) is quasi-invariant under transformations
(\ref{eq:infinitesimal}). Thus we get a (classical) dynamical
model, whose Lagrangian with higher order time derivatives has the
$\widehat{NH}$ algebra as its (quasi-)symmetry algebra.

\section{first order Lagrangian, Hamiltonian formalism and the extended phase space}\label{sec:1st-order}
\subsection{first order Lagrangian, Hamiltonian formalism}
Introducing $y_i=\dot{x_i}$ as independent coordinates, one can put
the Lagrangian (\ref{eq:Lagrangian}) into its first order form as
follows:
\begin{eqnarray}\label{eq:first-order}
L=p_i(\dot{x_i}-y_i)+q_i
\dot{y_i}-\frac{1}{2\kappa}{q_i}^2+\frac{\kappa}{2R^2} {y_i}^2
\end{eqnarray}
which can also be derived from the Lagrangian in Ref.\cite{GL}.
Using the Faddev-Jackiw procedure \cite{Faddeev,Jackiw} one obtains
the following nonvanishing Poisson brackets:
\begin{eqnarray}\label{eq:pb}
\{x_i, p_j\}=\delta_{ij},\qquad\{y_i,q_j\}=\delta_{ij}
\end{eqnarray}
The Hamiltonian which follows from (\ref{eq:first-order}) has the
form
\begin{eqnarray}\label{eq:Hamiltonian}
H=p_iy_i+\frac{1}{2\kappa} {q_i}^2-\frac{\kappa}{2R^2}{y_i}^2
\end{eqnarray}
The equations of motion which can be obtained from Euler-Lagrange
equations or Hamilton equations with
(\ref{eq:pb}-\ref{eq:Hamiltonian}), besides the equations
$y_i=\dot{x_i}$, are
\begin{eqnarray}
\dot{p_i}=0,\qquad\dot{y_i}=\frac{q_i}{\kappa},\qquad
p_i+\dot{q_i}-\frac{\kappa}{R^2}y_i=0
\end{eqnarray}
We take the infinitesimal $\widehat{NH}$ transformations of the
variables as
\begin{eqnarray}\label{eq:timein}
\delta t&=&\delta a \nonumber\\
\delta x_i&=&{(\delta O)_i}^jx_j+\cosh \frac{t}{R}\delta
a_i+R\sinh\frac{t}{R}\delta v_i+2R^2(\cosh\frac{t}{R}-1)\delta
b_i+y_i\delta a\nonumber\\
\delta y_i&=&{(\delta O)_i}^jy_j+\frac{1}{R}\sinh \frac{t}{R}\delta
a_i+\cosh\frac{t}{R}\delta v_i+2R\sinh\frac{t}{R}\delta
b_i+\frac{q_i}{\kappa}\delta a\nonumber\\
\delta q_i&=&{(\delta O)_i}^jq_j+\frac {\kappa}{R^2}\cosh
\frac{t}{R}\delta a_i+\frac{\kappa}{R}\sinh\frac{t}{R}\delta
v_i+2\kappa \cosh\frac{t}{R}\delta b_i-p_i\delta a+\frac
{\kappa}{R^2}y_i\delta
a\nonumber\\
\delta p_i&=&{(\delta O)_i}^jp_j
\end{eqnarray}
We should note that these transformations leave the Lagrangian
(\ref{eq:first-order}) invariant. Indeed, performing the
transformations (\ref{eq:timein}), one obtains
\begin{eqnarray}
\delta L =\frac{d}{dt}(\frac{\kappa}{R^2}y_i\cosh\frac{t}{R}\delta a_i
+\frac{\kappa}{R}y_i\sinh\frac{t}{R}\delta v_i+2\kappa y_i\cosh\frac{t}{R}\delta b_i-y_ip_i\delta a
+\frac{\kappa}{R^2}y_i^2\delta a)
\end{eqnarray}
which is a total derivative. We can also check that the
infinitesimal transformations (\ref{eq:timein}) in phase space are
canonical transformations. In fact, the transformed canonical
coordinates and canonical momenta are
\begin{eqnarray}
x_i'&=&x_i+{(\delta O)_i}^j x_j+\cosh \frac{t}{R}\delta
a_i+R\sinh\frac{t}{R}\delta v_i+2R^2(\cosh\frac{t}{R}-1)\delta
b_i+y_i\delta a\nonumber\\
y_i'&=&y_i+{(\delta O)_i}^j y_j+\frac{1}{R}\sinh \frac{t}{R}\delta
a_i+\cosh\frac{t}{R}\delta v_i+2R\sinh\frac{t}{R}\delta
b_i+\frac{q_i}{\kappa}\delta a\nonumber\\
q_i'&=&q_i+{(\delta O)_i}^j q_j+\frac {\kappa}{R^2}\cosh
\frac{t}{R}\delta a_i+\frac{\kappa}{R}\sinh\frac{t}{R}\delta
v_i+2\kappa \cosh\frac{t}{R}\delta b_i-p_i\delta a+\frac
{\kappa}{R^2}y_i\delta
a\nonumber\\
p_i'&=&p_i+{(\delta O)_i}^j p_j
\end{eqnarray}
and it is easy to show that the Poisson brackets of canonical
variables, up to first order of the infinitesimal parameters, are
\begin{eqnarray}\label{eq:canonical}
\{x_i',p_j'\}=\delta_{ij},\qquad\{y_i',q_j'\}=\delta_{ij}
\end{eqnarray}
with the others vanishing.

We can also get the Noether charges by the Noether theorem
\cite{Desloge} corresponding to the Lagrangian
(\ref{eq:first-order}). When the time $t$ is fixed, the
$\widehat{NH}$ transformations become
\begin{eqnarray}
\delta x_i&=&{(\delta O)_i}^jx_j+\cosh \frac{t}{R}\delta
a_i+R\sinh\frac{t}{R}\delta v_i+2R^2(\cosh\frac{t}{R}-1)\delta
b_i\nonumber\\
\delta y_i&=&{(\delta O)_i}^jy_j+\frac{1}{R}\sinh \frac{t}{R}\delta
a_i+\cosh\frac{t}{R}\delta v_i+2R\sinh\frac{t}{R}\delta
b_i\nonumber\\
\delta q_i&=&{(\delta O)_i}^jq_j+\frac {\kappa}{R^2}\cosh
\frac{t}{R}\delta a_i+\frac{\kappa}{R}\sinh\frac{t}{R}\delta
v_i+2\kappa
\cosh\frac{t}{R}\delta b_i\nonumber\\
\delta p_i&=&{(\delta O)_i}^jp_j
\end{eqnarray}
Using the Noether theorem we can obtain the Noether charges as
follow:
\begin{eqnarray}
P_i&=&p_i\cosh\frac{t}{R}+q_i\frac{1}{R}\sinh\frac{t}{R}-\frac{\kappa}{R^2}y_i\cosh\frac{t}{R}\nonumber\\
K_i&=&p_iR\sinh\frac{t}{R}+q_i\cosh\frac{t}{R}-\frac{\kappa}{R}y_i\sinh\frac{t}{R}\nonumber\\
F_i&=&2p_iR^2(\cosh\frac{t}{R}-1)+2q_iR\sinh\frac{t}{R}-2\kappa
y_i\cosh\frac{t}{R}
\end{eqnarray}
It is easy to see that the Noether charges become the same as that
in the acceleration-extended Galileo case, when
$R\rightarrow\infty$.

\subsection{Sch\"{o}dinger equation of the system and its invariance}
We describe the quantization of classical system
(\ref{eq:Hamiltonian}) by the method of geometric quantization
here. It is known that the classical system is described by the
Poisson algebra of observables defined on the phase space of the
system. If we quantized the classical system, we can obtain the
corresponding quantum states on the Hilbert space. The relation of
Poisson brackets of the classical observables and commutators of
Hermitian operators is:
\begin{eqnarray}\label{eq:quantize}
[\hat{f}_1,\hat{f}_2]=i\hbar\{f_1,f_2\}
\end{eqnarray}
where $f_1, f_2$ are the  classical observables   and $\hat{f}_1,
\hat{f}_2$ their quantum counterparts. Here,  corresponding to the
brackets (\ref{eq:pb}), the  commutators are:
\begin{eqnarray}
[\hat{x}_i,\hat{p}_j]=i\hbar\delta_{ij},\qquad [\hat{y}_i,\hat{q}_j]=i\hbar\delta_{ij}
\end{eqnarray}

For a classical system whose degrees of freedom is $n$, generalized
coordinates are $Q_i$ and generalized  momenta are $P_i$
($i=1,\cdots,n$).  The symplectic 2-form on the phase space $M$ is
\begin{eqnarray}
\omega=dP_i\wedge dQ_i
\end{eqnarray}
Now, in our case, the symplectic 2-form is
\begin{eqnarray}
\omega=dp_i\wedge dx_i+dq_i\wedge dy_i
\end{eqnarray}
Corresponding to every observable as a function $f$ on the phase
space $M$, we can define a vector field $X_f$ satisfying
\begin{eqnarray}
\imath_{X_f}\omega+df=0
\end{eqnarray}

We should pre-quantize the observables first, i.e. give the
Hermitian operator $\hat{f} $ corresponding to every observable $f$,
satisfying (\ref{eq:quantize}). The Hermitian operator can be
given by the following equation \cite{S}:
\begin{eqnarray}
\hat{f}=-i\hbar(X_f-\frac{i}{\hbar}\imath_{X_f}\theta)+f
\end{eqnarray}
where $\theta$ is the symplectic potential, i.e. the 1-form
satisfying $\omega=d\theta$. In our case, the symplectic potential
can be simply taken as
\begin{eqnarray}
\theta=p_i dx^i+q_i dy^i
\end{eqnarray}
Then we should polarize the Hermitian operators. Because our case is
relatively simple, the Hermitian operators obtained after
polarization are just the Hermitian operators on the Hilbert space,
corresponding to the classical observables.

Thus we can obtain the Hermitian operators on the Hilbert space
corresponding to the observables in NH space-time in this way. The
operators are:
\begin{eqnarray}
\hat{x}_i=x_i,\qquad\hat{p}_i=-i\hbar\frac{\partial}{\partial
x_i}\nonumber\\
\hat{y}_i=y_i,\qquad\hat{q}_i=-i\hbar\frac{\partial}{\partial y_i}
\end{eqnarray}
So the Schr\"odinger equation of the system can be obtained as
\begin{eqnarray}\label{eq:schrodinger}
(-\frac{\hbar^2}{2\kappa}\frac{\partial}{\partial
y_i}\frac{\partial}{\partial y_i}-i\hbar y_i\frac{\partial}{\partial
x_i}-\frac{\kappa}{2R^2}{y_i}^2)\varphi=i\hbar\frac{\partial}{\partial
t}\varphi
\end{eqnarray}

Now we can verify that the equation (\ref{eq:schrodinger}) is
invariant under infinitesimal transformations (\ref{eq:timein}).
First, we get the Hermitian operators corresponding to the
transformed observables, again using the geometric quantization:
\begin{eqnarray}
\hat{x}_i'&=&x_i+{(\delta O)_i}^jx_j+\cosh\frac{t}{R}\delta
a_i+R\sinh\frac{t}{R}\delta v_i+2R^2(\cosh\frac{t}{R}-1)\delta
b_i=x'_i\nonumber\\
\hat{y}_i'&=&y_i+{(\delta O)_i}^j y_j+\frac {1}{R}\sinh
\frac{t}{R}\delta a_i+\cosh\frac{t}{R}\delta v_i+2R\sinh
\frac{t}{R}\delta b_i-i\hbar\frac{\delta
a}{\kappa}\frac{\partial}{\partial
y_i}\nonumber\\
\hat{p}_i'&=&-i\hbar(\frac{\partial}{\partial x_i}+{(\delta
O)_i}^j\frac{\partial}{\partial
x_j})\\
\hat{q}_i'&=&-i\hbar(\frac{\partial}{\partial y_i}+{(\delta
O)_i}^j\frac{\partial}{\partial y_j})-i\hbar\delta a
\frac{\partial}{\partial
x_i}+\frac{\kappa}{R^2}\cosh\frac{t}{R}\delta
a_i+\frac{\kappa}{R}\sinh\frac{t}{R}\delta
v_i+2\kappa\cosh\frac{t}{R}\delta b_i+\frac{\kappa}{R^2} y_i\delta
a\nonumber
\end{eqnarray}
It is easy to check that the above transformed operators still
satisfy the standard commutators
\begin{equation}
[\hat{x}_i',\hat{p}_j']=i\hbar\delta_{ij},\qquad [\hat{y}_i',\hat{q}_j']=i\hbar\delta_{ij}
\end{equation}
as already indicated by equations (\ref{eq:canonical}) and
(\ref{eq:quantize}). As is well known, in this case the
transformations act on the Hilbert space as similarity
transformations. Under certain Hermitian condition, these
transformations become unitary ones.

Then, for simplicity, we consider rotations, spatial translations,
time translation, accelerations and boosts one by one. First, it is
obvious that equation (\ref{eq:schrodinger}) is invariant under
rotations if the wave function $\varphi(x,y,t)$ is invariant.
Second, the spatial translations  are
\begin{eqnarray}\label{eq:spatial}
\hat{x}_i'&=&x_i+\cosh \frac{t}{R}\delta a_i\nonumber\\
\hat{y}_i'&=&y_i+\frac {1}{R}\sinh\frac{t}{R}\delta a_i\nonumber\\
\hat{p}_i'&=&-i\hbar\frac{\partial}{\partial x_i}\nonumber\\
\hat{q}_i'&=&-i\hbar\frac{\partial}{\partial
y_i}+\frac{\kappa}{R^2}\cosh\frac{t}{R}\delta a_i\nonumber\\
t'&=&t
\end{eqnarray}
From the chain rule of derivatives, we can obtain the transformed
time derivative:
\begin{eqnarray}\label{eq:timetranslation}
\frac{\partial}{\partial t'}&=&\frac{\partial}{\partial
t}\frac{\partial t}{\partial t'}+\frac{\partial}{\partial x_i
'}\frac{\partial x_i'}{\partial t'}+\frac{\partial}{\partial
y_i'}\frac{\partial y_i'}{\partial t'}+\frac{\partial}{\partial
q_i'}\frac{\partial q_i'}{\partial t'}\nonumber\\
& =&\frac{\partial}{\partial t}+\frac{\delta
a_i}{R}\sinh\frac{t}{R}\frac{\partial}{\partial x_i}+\frac{\delta
a_i}{R^2}\cosh \frac{t}{R}\frac{\partial}{\partial y_i}+\frac{\delta
a_i}{R^3}\kappa\sinh \frac{t}{R}\frac{\partial}{\partial q_i}
\end{eqnarray}
If we substitute equations (\ref{eq:spatial}) and
(\ref{eq:timetranslation}) into equations (\ref{eq:Hamiltonian}) and
(\ref{eq:schrodinger}) with every quantity replaced by the
corresponding primed one, omitting the higher order terms and
because of $\frac{\partial}{\partial q_i'}=\frac{1}{i\hbar}y_i'$, we
will obtain the following equation:
\begin{eqnarray}\label{eq:sschrodinger}
(-\frac{\hbar^2}{2\kappa}\frac{\partial}{\partial
y_i}\frac{\partial}{\partial y_i}-i\hbar y_i\frac{\partial}{\partial
x_i}-\frac{\kappa}{2R^2}{y_i}^2)\varphi'=i\hbar\frac{\partial}{\partial
t}\varphi'
\end{eqnarray}
It is now obvious that equation (\ref{eq:schrodinger}) is invariant
under spatial translations, if the the wave function
$\varphi(x,y,t)$ is itself invariant. Similarly, we can also show
that the Schr\"odinger equation is invariant under time translation,
boosts and acceleration transformations.

\section{exotic conformal Newton-Hooke algebra and its dynamical realization in $D=2+1$ dimension}
\subsection{exotic conformal Newton-Hooke algebra}
To obtain the conformal NH algebra, one has to add to the
acceleration-extended NH algebra
(\ref{eq:NH}-\ref{eq:last},\ref{eq:acceleration}) another two
generators, dilatation $D$ and expansion $K$, which together with
the Hamiltonian form a subalgebra:
\begin{eqnarray}
[D,H]=-H-\frac{1}{2R^2}K,\qquad [K,H]=-2D,\qquad [D,K]=K
\end{eqnarray}
The generators $D$ and $K$ are scalars:
\begin{eqnarray}
[D,J]=[K,J]=0
\end{eqnarray}
Here, $J$ is the rotation generator defined as $J_{ij}=\varepsilon_{ij}J$ in
$D=2+1$, with $\varepsilon_{ij}$ the standard antisymmetric tensor:
$\varepsilon_{12}=-\varepsilon_{21}=1$. $D$ and $K$ also satisfy
\begin{eqnarray}
&&[D,P_i]=-P_i+\frac{1}{2R^2}F_i,\qquad [D,K_i]=0,\qquad
[D,F_i]=F_i\\
&&[K,P_i]=-2K_i,\qquad [K,K_i]=-F_i,\qquad [K,F_i]=0.
\end{eqnarray}
One can show that the realization of the above Lie algebra on the
$D=2+1$ nonrelativistic space-time is given in terms of differential
operators as
\begin{eqnarray}
H&=&\partial_t,\qquad P_i=-\cosh\frac{t}{R}\partial_i,\qquad K_i=-\sinh\frac{t}{R}\partial_i\nonumber\\
F_i&=&-2R^2(\cosh\frac{t}{R}-1)\partial_i,\qquad D=R\sinh\frac{t}{R}\partial_t+\cosh\frac{t}{R}x_i\partial_i\nonumber\\
K&=&2R^2(\cosh\frac{t}{R}-1)\partial_t+2R\sinh\frac{t}{R}x_i\partial_i,\qquad
J=-\varepsilon_{ij}x_i\partial_j
\end{eqnarray}
It is easy to see that the conformal NH algebra goes back to conformal Galileo
algebra (see \cite{Lukierski}) in the limit of
$R\rightarrow\infty$.

It is easy to work out the spacetime infinitesimal transformations
generated by the conformal NH algebra:
\begin{eqnarray}\label{eq:conformal}
\delta x_i&=&-\cosh\frac{t}{R} \delta a_i-\sinh\frac{t}{R}\delta
v_i-2R^2(\cosh\frac{t}{R}-1)\delta b_i+\cosh\frac{t}{R} x_i\delta
d+2R\sinh\frac{t}{R}x_i\delta e+\delta
\alpha\varepsilon_{ij}x_j\nonumber\\
\delta t&=&\delta a+R\sinh\frac{t}{R}\delta
d+2R(\cosh\frac{t}{R}-1)\delta e
\end{eqnarray}
where $\delta d$ is the infinitesimal parameter for the dilatation
$D$, and $\delta e$ that for the expansion $K$.

One can extend the conformal NH algebra by introducing an exotic
 central element $\Theta$ in (2+1) dimension. This element is
introduced into the Lie bracket for two NH boosts \cite{L}:
\begin{eqnarray}\label{eq:boost}
[K_i,K_j]=\Theta\varepsilon_{ij}
\end{eqnarray}
As a consequence the Lie-bracket $[P_i,K_j]$ becomes also
nonvanishing:
\begin{eqnarray}\label{eq:pk}
[P_i,K_j]=-2\Theta\varepsilon_{ij}.
\end{eqnarray}

The conformal NH algebra with the modified relations
(\ref{eq:boost},\ref{eq:pk}) will be called exotic conformal
Newton-Hooke algebra hereafter.

\subsection{the dynamical model of the conformal Newton-Hooke symmetry}

As in the case of the acceleration-extended NH algebra (see
\cite{Stichel}), we should also find a Lagrangian which is
quasi-invariant under the conformal NH transformation
(\ref{eq:conformal}) and reduces to that in the Galilei case in the
limit of $R\rightarrow\infty$, i.e. \cite{Zakrzewski}
\begin{eqnarray}
L|_{R\rightarrow\infty}=-\frac{\Theta}{2}\varepsilon_{ij}\dot x_i
\ddot x_j
\end{eqnarray}
 Again, we can take the following substitution
\begin{eqnarray}
\varepsilon_{ij}\dot x_i\ddot{x_i} \rightarrow \varepsilon_{ij}\dot
x_i\ddot{x_i}+\frac {1}{R^2}x_i\dot x_i
\end{eqnarray}
to get the dynamical model of the conformal NH symmetry. So we
obtain the higher order Lagrangian of the conformal NH symmetry as
follows
\begin{eqnarray}\label{eq:fo}
L=-\frac{\Theta}{2}\varepsilon_{ij}\dot x_i\ddot{x_i}+\frac
{1}{R^2}x_i\dot x_i
\end{eqnarray}
 We can show that the Lagrangian is
quasi-invariant under NH conformal transformations
(\ref{eq:conformal}). Indeed, performing the Lagrangian(\ref{eq:fo})
one obtains
\begin{eqnarray}
\delta L&=&\frac{d}{dt}[\frac{\Theta}{2}\varepsilon_{ij}(\dot
x_j\frac{1}{R}\sinh\frac{t}{R}+\frac{1}{R^2}x_j\cosh\frac{t}{R})\delta
a_i+\frac{\Theta}{2}\varepsilon_{ij}(\dot
x_j\frac{1}{R}\cosh\frac{t}{R}+\frac{1}{R^2}x_j\sinh\frac{t}{R})\delta
v_i\\& & +\Theta\varepsilon_{ij}(\dot x_j
R\sinh\frac{t}{R}-x_j\cosh\frac{t}{R}-x_j)\delta
b_i-\frac{\Theta}{2}\varepsilon_{ij}\dot{x_j}{x_i}\frac{1}{R}\sinh\frac{t}{R}
\delta d-\Theta\varepsilon_{ij}\dot{x_j}x_i\cosh\frac{t}{R}\delta
e]\nonumber
\end{eqnarray}
Introducing $y_i=\dot x_i$ as an independent coordinate the
Lagrangian (\ref{eq:fo}) can be put into the following first order
form
\begin{eqnarray}\label{eq:l}
L=P_i(\dot x_i-y_i)-\frac{\Theta}{2}\varepsilon_{ij}(y_i\dot
y_j+\frac{1}{R^2}x_i y_i)
\end{eqnarray}
It can be checked that the first order form of the Lagrangian
(\ref{eq:l}) becomes that in Galileo case in the limit of
$R\rightarrow\infty$, see \cite{Zakrzewski}. Then we can obtain the
Hamiltonian following from the Lagrangian (\ref{eq:l}):
\begin{eqnarray}
H=P_iy_i+\frac{\Theta}{2R^2}\varepsilon_{ij}x_iy_j
\end{eqnarray}
One can also obtain the following Poisson brackets from (\ref{eq:l})
due to the Faddeev-Jackiw procedure \cite{Faddeev,Jackiw}:
\begin{eqnarray}
\{x_i,P_j\}=\delta_{ij},\qquad\{y_i,y_j\}=\frac{\varepsilon_{ij}}{\Theta}
\end{eqnarray}
The equations of motion, which can be obtained from the
Euler-Lagrange equations or the Hamilton equations, are:
\begin{eqnarray}
y_i=\dot x_i,\qquad\dot
y_i-\frac{\varepsilon_{ij}}{\Theta}P_j+\frac{1}{2R^2}\varepsilon_{ij}\dot
x_j=0,\qquad\dot P_i+\frac{\Theta}{2R^2}\varepsilon_{ij}y_j=0
\end{eqnarray}

\section{conclusion}
Following Ref.\cite{Stichel}, we mainly discuss the acceleration-extended
Newton-Hooke algebra and its dynamical realization in
arbitrary dimension in this paper. We first recall the NH algebra,
and then extend it by adding acceleration generators to it. The
central charges are also introduced. We present the Lagrangian with
higher order time derivatives that is quasi-invariant under the
$\widehat{NH}$ transformations. After that, we obtain the quantities
on the phase space using the first order form of the Lagrangian, and
show that the $\widehat{NH}$ transformations on the phase space are
canonical transformations. The Noether charges corresponding to the
$\widehat{NH}$ transformations are obtained from the Noether
theorem. The Schr\"{o}dinger equation is obtained in standard way,
and by geometric quantization we also show that it is invariant
under the $\widehat{NH}$ transformations, which now act on the
corresponding Hilbert space as unitary transformations. Finally, we
discuss the exotic conformal NH algebra and its dynamical
realization briefly.

It is easy to see that all the results in this paper go back to
those in the Galileo case in the limit of $R\rightarrow\infty$. On
the other hand, although we give only the explicit results on the NH
space-time contracted from the dS one, all the discussions in our
paper can be as well applied to the AdS case. The quantum dynamical
realization of the exotic conformal NH symmetry is left for future
work. Further investigations on some possible physical implications
of these results are even more interesting.

\begin{acknowledgments}
The authors would like to thank Prof. Xiaoning Wu for valuable
discussions. This work is partly supported by the National Natural
Science Foundation of China (Grant No. 10605005).
\end{acknowledgments}

\end{document}